\pdfoutput=1
\pdfpageattr{/Group << /S /Transparency /I true /CS /DeviceRGB>>}
\documentclass[aps,prl,reprint,groupedaddress,showpacs,amsmath]{revtex4-1}
\usepackage[T1]{fontenc}
\usepackage[utf8]{inputenc}

\usepackage{isotope}
\usepackage{xcolor}
\usepackage{graphicx}
\usepackage[load-configurations=abbreviations]{siunitx}
\usepackage{tikz}
\usepackage{booktabs}

\usepackage{txfonts}

\usepackage{hyperref}
\usepackage[poorman,capitalise]{cleveref}

\makeatletter
\DeclareUnicodeCharacter{014C}{\@tabacckludge=O} % LATIN CAPITAL LETTER O WITH MACRON
\DeclareUnicodeCharacter{014D}{\@tabacckludge=o} % LATIN SMALL LETTER O WITH MACRON
\makeatother

\ifpdf\else\renewcommand{\href}[2]{#2}\fi\newcommand{\includegr}[1]{\includegraphics{#1}}

\DeclareSIUnit\keV{\kilo\electronvolt}
\DeclareSIUnit\MeV{\mega\electronvolt}
\DeclareSIUnit\MeVc{\MeV\per\text{\ensuremath{c}}}
\DeclareSIUnit\fm{\femto\meter}
\DeclareSIUnit\quadrupoleunit{\elementarycharge\tothe2\fm\tothe4}
\DeclareSIUnit\Msol{\ensuremath{M_\odot}}

\newcommand{\sixdash}[1][]{\tikz[baseline=-2.5pt]{\draw[line width=1.5pt, dash pattern=on 1.98pt off 1.98pt, #1] (0,0) -- +(9.9pt,0pt);}}
\newcommand{\sevendash}[1][]{\tikz[baseline=-2.5pt]{\draw[line width=1.5pt, #1] (0,0) -- +(9.9pt,0pt);}}
\newcommand{\baredash}[1][]{\tikz[baseline=-2.5pt]{\draw[line width=1.5pt, dash pattern=on 3.3pt off 3.3pt, #1] (0,0) -- +(9.9pt,0pt);}}
\newcommand{\evolveddash}[1][]{\tikz[baseline=-2.5pt]{\draw[line width=1.5pt,#1] (0,0) -- +(8pt,0pt);}}

\newcommand{\alphaNN}{\ensuremath{\alpha_{N}}}
\newcommand{\alphaYN}{\ensuremath{\alpha_{Y}}}
\newcommand{\Nmax}{\ensuremath{N_\text{max}}}

\newcommand{\doubletoprule}{\toprule[\lightrulewidth]\toprule[\lightrulewidth]}

\begin{document}

%\preprint{}

\title{Induced Hyperon-Nucleon-Nucleon Interactions and the Hyperon Puzzle}

\author{Roland Wirth}
\email{roland.wirth@physik.tu-darmstadt.de}

\author{Robert Roth}
\email{robert.roth@physik.tu-darmstadt.de}
\affiliation{Institut f\"ur Kernphysik, Technische Universit\"at Darmstadt, Schlossgartenstr. 2, 64289 Darmstadt, Germany}

\date{\today}

\begin{abstract}
We present the first \emph{ab initio} calculations for $p$\nobreakdash-shell hypernuclei including hyperon--nucleon--nucleon (YNN) contributions induced by a Similarity Renormalization Group transformation of the initial hyperon--nucleon interaction.
The transformation including the YNN terms conserves the spectrum of the Hamiltonian while drastically improving model-space convergence of the Importance-Truncated No-Core Shell Model, allowing a precise extraction of binding and excitation energies.
Results using a hyperon--nucleon interaction at leading order in chiral effective field theory for lower- to mid-$p$-shell hypernuclei show a good reproduction of experimental excitation energies while hyperon separation energies are typically overestimated.
The induced YNN contributions are strongly repulsive and we show that they are related to a decoupling of the $\Sigma$ hyperons from the hypernuclear system, i.e., a suppression of the $\Lambda$--$\Sigma$ conversion terms in the Hamiltonian.
This is linked to the so-called hyperon puzzle in neutron-star physics and provides a basic mechanism for the explanation of strong  $\Lambda$NN three-baryon forces.
\end{abstract}

%\keywords{}

\maketitle

The exploration of the structure of hypernuclei is a long-standing focus of both experimental and theoretical efforts.
From the first detection of a hypernucleus \cite{Danysz1953,Davis1986,*Davis2005} to recent discoveries, like the ``strange alpha'' \isotope[6][\Lambda\Lambda]{He} \cite{Takahashi2001} or the extremely neutron-rich \isotope[6][\Lambda]{H} \cite{Agnello2012}, experiments have provided a wealth of data, not only on ground-state energies but also on spectroscopic observables like excitation energies \cite{May1983,Akikawa2002,Hashimoto2006,Yamamoto2015,Gogami2016} or transition strengths \cite{Tanida2001}.
New experiments, e.g.\ at J-Parc, JLab or FAIR, will continue to expand our knowledge about hypernuclei throughout the hypernuclear chart \cite{Gal2016}.
This program is accompanied by a long history of phenomenological models, most notably, the shell model for p- and sd-shell hypernuclei \cite{Gal1971,*Gal1972,*Gal1978,Millener2008,*Millener2010,*Millener2011,*Millener2012,Gal2013}, cluster models \cite{Motoba1983,Motoba1985,Hiyama2009,Hiyama2012,Hiyama2014}, various mean-field models \cite{Vretenar1998,Glendenning1993,Vidana2001,Guleria2012}, or recent Monte Carlo calculations with simplified phenomenological interactions \cite{Lonardoni2013,Lonardoni2014}.
However, for a fundamental understanding of the underlying interactions and the full dynamics of hypernuclei we need \emph{ab initio} methods that eliminate model dependencies in the solution of the many-baryon problem.
For a long time, these were only available for systems of up to four nucleons \cite{Nemura2002,Nogga2002,Haidenbauer2007,Nogga2013,Gazda2016a}.

In order to advance the understanding of hypernuclei and hyperon--nucleon interactions from first principles, we recently presented the first \emph{ab initio} calculations of $p$-shell hypernuclei with realistic (chiral) baryon--baryon interactions \cite{Wirth2014}, performed in an Importance-Truncated No-Core Shell Model (IT-NCSM) framework \cite{Roth2009}.
These calculations, however, suffered from slow convergence of energies with respect to the NCSM model-space size.
For non-strange nuclear systems it is common practice to improve convergence by preconditioning the Hamiltonian, e.g.\ via a Similarity Renormalization Group (SRG) transformation \cite{Gazek1993,Wegner1994,*Wegner2000,White2002,Bogner2007,Bogner2010,Roth2011} that prediagonalizes the Hamiltonian in momentum space.
This transformation induces three- and higher many-body forces even if the initial Hamiltonian contains only two-body terms.
In light non-strange nuclei the induced three-nucleon terms are of moderate size and can be explicitly included into the IT-NCSM calculation by performing the SRG transformation in three-body space.
When transforming the hyperon--nucleon (YN) interaction the induced terms are much stronger.
We observe \SI{4.1}{\MeV} overbinding relative to a $\Lambda$ separation energy of $B_\Lambda=\SI{6.4}{\MeV}$ for \isotope[7][\Lambda]{Li} (cf.\ first and third column of the third row of \cref{tab:blambda}).
In our previous work we hence worked with unevolved YN interactions, which caused the aforementioned slow convergence, especially of the absolute binding energies.

Hyperons and their interactions with nucleons also have an impact on neutron-star physics.
When describing the structure of neutron stars assuming purely nucleonic matter in $\beta$ equilibrium, their interior reaches densities and nucleon chemical potentials that would make the appearance of hyperons energetically favorable.
However, if one includes hyperons into the description, assuming typical phenomenological YN interactions, the equation of state softens and the maximum neutron-star mass drops significantly \cite{Vidana2015,Lonardoni2015,Gal2016}.
As a consequence, these equations of state cannot support neutron stars with about \SI{2}{\Msol} that were observed recently \cite{Demorest2010,Antoniadis2013}.
This is the so-called hyperon puzzle in neutron-star physics.

In this Letter we present calculations in the $p$ shell including not only chiral and induced three-nucleon (3N), but also induced hyperon--nucleon--nucleon (YNN) interactions.
We show that it is the suppression of the $\Lambda$--$\Sigma$ conversion that drives the induced YNN terms and that these terms are mainly of three-body nature.
This observation leads to new insights into the hyperon puzzle.

\paragraph{Similarity Renormalization Group.}
The SRG transformation \cite{Gazek1993,Wegner1994,*Wegner2000,White2002,Bogner2007,Bogner2010,Jurgenson2009,Roth2011} is a continuous unitary transformation $H(\alpha)=U^\dag(\alpha) H(0) U(\alpha)$ of the hypernuclear Hamiltonian
$H = H(0) = T_\text{int} + V_\text{NN} + V_\text{3N} + V_\text{YN} + \Delta M$
consisting of kinetic energy and NN and 3N interaction terms along with a YN interaction and a mass term accounting for the different rest masses of the $\Lambda$ and $\Sigma$ hyperons.
Coulomb interactions among charged baryons are contained in the two-body terms.
The transformation is governed by the flow equation $\partial_\alpha H(\alpha) = [\eta(\alpha),H(\alpha)]$ with flow parameter $\alpha$.
The anti-hermitian generator $\eta(\alpha)$ can be chosen freely in order to achieve a desired behavior.
Here, we adopt the common choice $\eta(\alpha) = m_N^2 [T_\text{int}, H(\alpha)]$ that drives the Hamiltonian to a band-diagonal form in momentum space.
The flow equation can be evaluated on a sufficiently large basis set (e.g.\ consisting of harmonic-oscillator states) and solved as an ordinary matrix differential equation using standard numerical methods.

In order to capture the induced YNN terms we have to evaluate the flow equation in a three-body basis, which we construct from HO wave functions with respect to three-body Jacobi coordinates.
The resulting evolved Hamiltonian contains a mixture of two- and three-body terms that must be disentangled because they scale differently in a many-body calculation.
This is achieved by subtracting the Hamiltonian evolved in two-body space.
The resulting YNN interaction can then be used in a hypernuclear IT-NCSM framework \cite{Wirth2014}.
This procedure is well established for NN and 3N interactions \cite{Jurgenson2009,Roth2014}, but here we present the first calculations with induced YNN interactions.

\paragraph{Results for $p$-shell hypernuclei.}
\begin{figure}
  \includegr{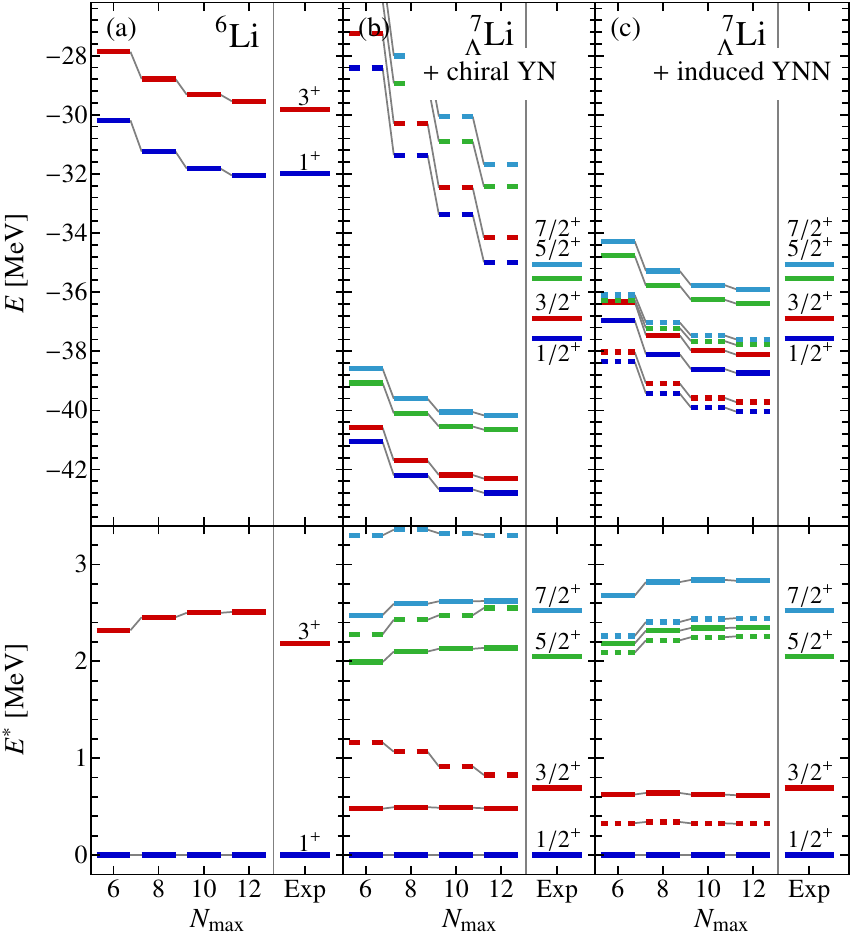}
  \caption{\label{fig:lli7}(color online)
    Absolute and excitation energies of \isotope[7][\Lambda]{Li} (a) Nucleonic parent absolute and excitation energies, (b) hypernucleus with bare (\baredash) and SRG-evolved (\evolveddash) YN interaction (\SI{700}{\MeVc} cutoff, evolved to $\alpha_Y=\SI{0.08}{\fm\tothe4}$), (c) hypernucleus with added YNN terms for cutoffs \SI{700}{\MeVc} (\sevendash) and \SI{600}{\MeVc} (\sixdash).
    The calculations are carried out with an NN+3N interaction evolved to $\alpha_N=\SI{0.08}{\fm\tothe4}$ in a HO basis with $\hbar\Omega=\SI{20}{\MeV}$.
  }
\end{figure}

\begin{figure}
  \includegr{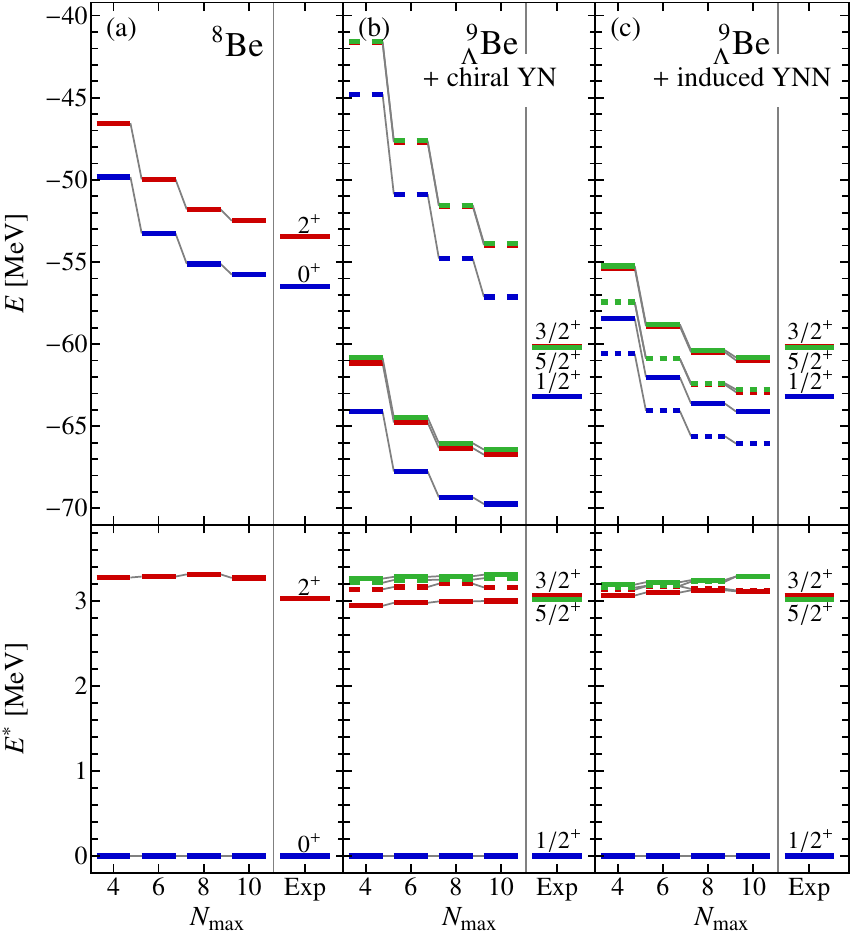}
  \caption{\label{fig:lbe9}(color online)
    Same as \cref{fig:lli7}, but for \isotope[9][\Lambda]{Be}.
  }
\end{figure}

\begin{figure}
  \includegr{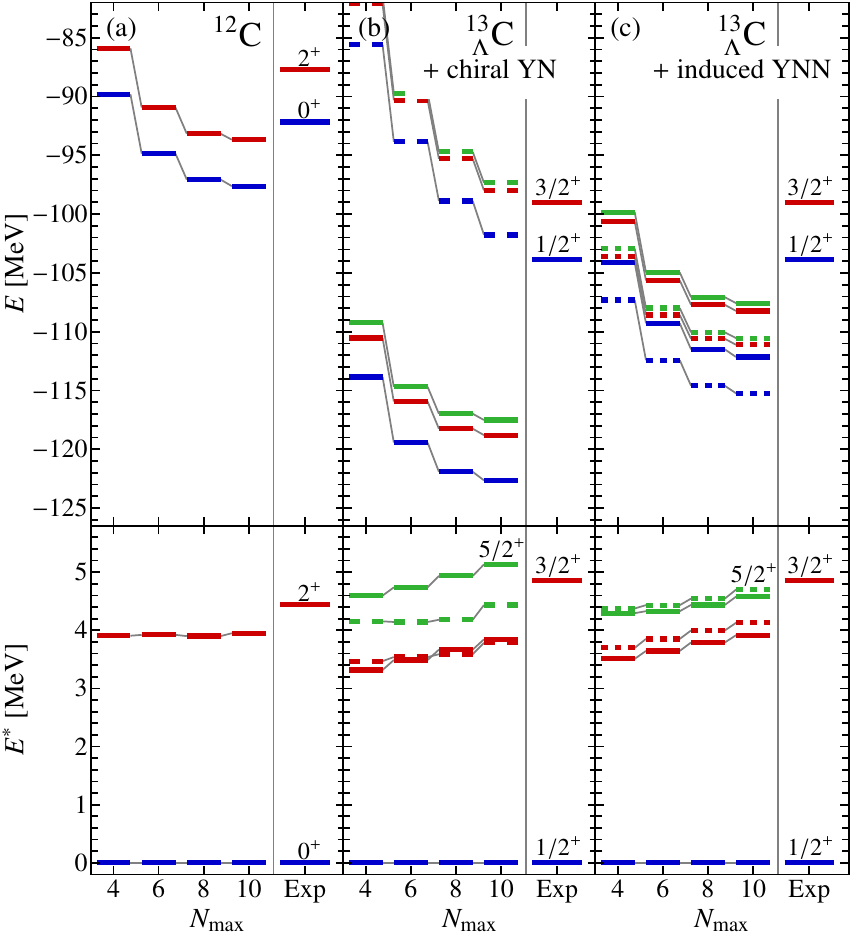}
  \caption{\label{fig:lc13}(color online)
    Same as \cref{fig:lli7}, but for \isotope[13][\Lambda]{C}.
  }
\end{figure}

To illustrate the effect of including SRG-induced YNN terms we show the absolute and excitation energies of low-lying states in \isotope[7][\Lambda]{Li}, \isotope[9][\Lambda]{Be} and \isotope[13][\Lambda]{C} in \cref{fig:lli7,fig:lbe9,fig:lc13} together with the energies of the non-strange parent nuclei (left-hand panels).
We are using the standard chiral NN interaction at next-to-next-to-next-to-leading order (N\textsuperscript3LO) \cite{Entem2003} and the chiral 3N interaction at N\textsuperscript2LO \cite{Navratil2007} (both with cutoff $\Lambda=\SI{500}{\MeVc}$) in conjunction with a chiral YN interaction at LO \cite{Polinder2006} (with cutoffs $\Lambda=\SI{700}{\MeVc}$ and $\Lambda=\SI{600}{\MeVc}$).
For all following calculations we perform a consistent SRG evolution of the NN and 3N interaction up to the three-nucleon level.
In previous \emph{ab initio} calculations for p-shell nuclei, we have shown the SRG-induced beyond-3N interactions are sufficiently small in the mass range considered here \cite{Roth2014}.

The center panels of \cref{fig:lli7,fig:lbe9,fig:lc13} compare the energy spectra of the single-strangeness hypernuclei obtained with the bare YN interaction and an SRG-evolved YN interaction at the two-baryon level with $\alpha_Y=\SI{0.08}{\fm\tothe4}$.
The SRG evolution of the YN interaction causes a large drop of the absolute energies of all states while convergence with respect to the model-space parameter \Nmax{} is greatly improved.
Excitation energies also show much faster convergence in \isotope[7][\Lambda]{Li} while the effect for the other isotopes is less dramatic.
The splittings of the excited state doublets in \isotope[9][\Lambda]{Be} and \isotope[13][\Lambda]{C} are increased.
The convergence patterns of the nucleonic parent and the hypernuclear states become very similar, which allows for a precise extraction of the $\Lambda$ separation energy with only a few hundred \si{\keV} uncertainty.
These separation energies are summarized in \cref{tab:blambda}.

In a next step, we include the SRG-induced YNN terms explicitly in our calculations---the results are shown in the right-hand panels of \cref{fig:lli7,fig:lbe9,fig:lc13}.
Evidently, the induced YNN terms counteract the drop of the absolute energies and shift them closer to the values extrapolated from the bare YN result.
Convergence patterns and excitation energies are barely affected, implying that the induced YNN terms act on all states in the same manner.
This is in accordance with the behavior of SRG-induced nucleonic 3N forces.
A notable effect is the increase of the doublet splittings in \isotope[7][\Lambda]{Li} compared to the case without induced YNN terms, while those in \isotope[9][\Lambda]{Be} and \isotope[13][\Lambda]{C} are reduced.
In conclusion, when one accounts for the sizable cutoff uncertainty in the YN interaction the excitation energies including the induced YNN terms are compatible with experimental data \cite{Hashimoto2006}; however, the hyperon is overbound significantly by \SIrange{20}{50}{\percent} \cite{Davis1986,*Davis2005}, depending on the YN interaction cutoff.
Overall, the YN interaction with cutoff $\Lambda=\SI{700}{\MeVc}$ provides a consistently better description of the hypernuclei under consideration.

\paragraph{Evolution of YNN terms.}
\begin{figure}[t]
  \includegr{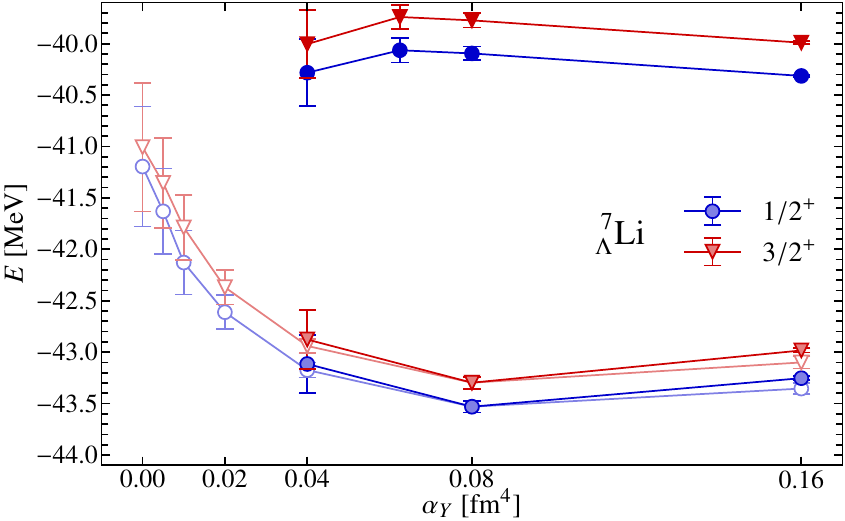}
  \caption{\label{fig:lli7-ynsrg}(color online)
  Extrapolated binding energy of the \isotope[7][\Lambda]{Li} ground-state doublet as a function of the SRG flow parameter $\alphaYN$ in the hyperon--nucleon sector, neglecting and including induced YNN contributions.
  Calculations without induced YNN terms for low values of $\alphaYN$ are carried out with nucleonic flow parameter fixed at $\alphaNN=\SI{0.08}{\fm\tothe4}$ (faded empty symbols), for higher values we take $\alphaNN=\alphaYN$ (half-filled symbols).
  The inclusion of the induced YNN terms (full symbols) restores flow-parameter independence within extrapolation uncertainties.
  The YN interaction cutoff is $\Lambda=\SI{600}{\MeVc}$, the HO frequency is $\hbar\Omega=\SI{20}{\MeV}$.}
\end{figure}
A detailed analysis of the emergence of SRG-induced YNN interactions is presented in \cref{fig:lli7-ynsrg}, which shows the extrapolated energies of the lowest two states of \isotope[7][\Lambda]{Li} as a function of the flow parameter $\alpha_Y$, with and without induced YNN terms.
The energies are obtained by simple exponential extrapolations of the calculated ground-state energies to the full Hilbert space and the quoted uncertainties include importance-threshold and model-space extrapolation uncertainties.
The absolute energies of the ground-state doublet show a strong $\alpha_Y$ dependence.
Adding the induced YNN terms practically removes the $\alpha_Y$ dependence and recovers the unevolved energies within extrapolation uncertainties.
From this we conclude that the induced terms are mainly of three-body nature and that the net contribution due to four- and higher many-body forces is small in these systems.
The induced YNN terms are surprisingly large, at $\alpha_Y \approx \SI{0.08}{\fm\tothe4}$ their inclusion changes the ground-state energy of \isotope[7][\Lambda]{Li} by about \SI{3.5}{\MeV}, which can be compared to a $\Lambda$ separation energy of about \SI{7.7}{\MeV}. %and an expectation value of the evolved YN potential of about 25 MeV.

\begin{figure}[t]
  \includegr{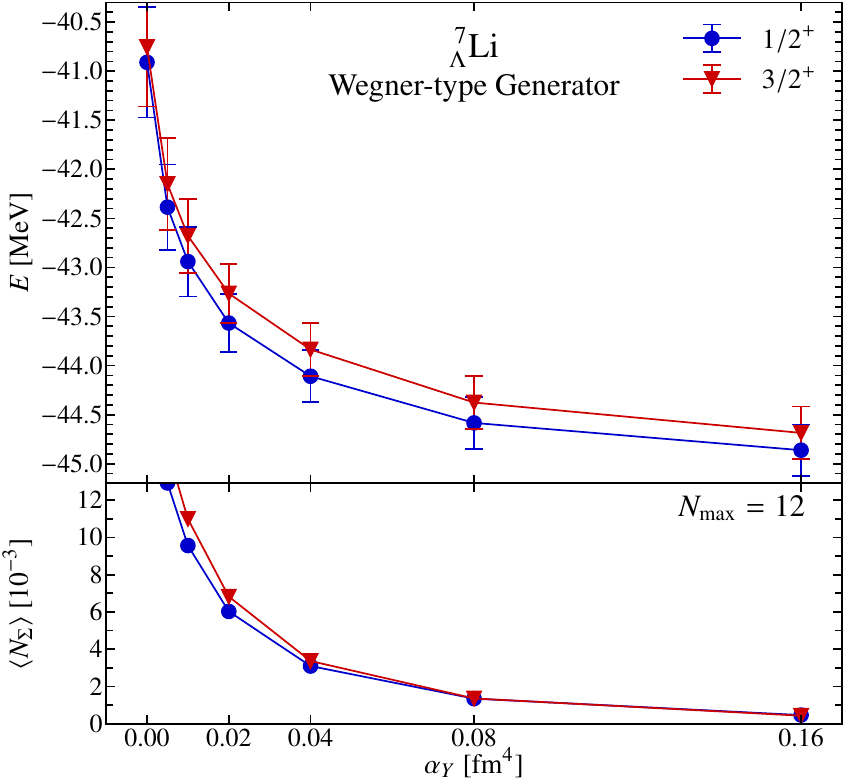}
  \caption{\label{fig:lli7-wegner}(color online)
  Extrapolated binding energy of the \isotope[7][\Lambda]{Li} ground-state doublet as a function of the SRG flow parameter $\alphaYN$ in the hyperon--nucleon sector using the Wegner-type generator discussed in the text.
  Calculations are carried out with the nucleonic flow parameter fixed at $\alphaNN=\SI{0.08}{\fm\tothe4}$, a HO frequency of $\hbar\Omega=\SI{20}{\MeV}$ and a YN interaction cutoff of $\Lambda=\SI{600}{\MeVc}$, the lower panel shows the expectation value of the $\Sigma$ number operator for the largest model space calculated.}
\end{figure}

As a possible origin of the large induced YNN contributions one might suspect a unique feature of the YN interaction, the conversion between $\Lambda$ and $\Sigma$ hyperons in the interaction with a nucleon \cite{Miyagawa1995,Akaishi2000,Hiyama2001b,Nogga2002}.
The SRG generator $\eta(\alpha)$ drives the Hamiltonian to a band-diagonal form, also causing a suppression of the $\Lambda$--$\Sigma$ conversion due to the presence of the mass and kinetic energy terms.
To further explore this hypothesis, we adopt Wegner's original formulation of the SRG \cite{Wegner1994,*Wegner2000} and design a generator that exclusively decouples the $\Lambda$ from the $\Sigma$ hyperons. The Wegner-type generator $\eta'(\alpha) = m_N^2 [H_{\Lambda\Sigma}(\alpha), H(\alpha)]$, where $H_{\Lambda\Sigma}$ contains only the $\Lambda$--$\Sigma$ conversion terms, does exactly this.
For large flow parameters $\alpha$, this results in a Hamiltonian whose low-lying spectrum is completely devoid of $\Sigma$ admixtures and, thus, corresponds to an effective $\Lambda$-only model.
The evolution of the energies of the ground and first-excited states in \isotope[7][\Lambda]{Li} without the induced YNN terms and the expectation values of the $\Sigma$ number operator are shown in \cref{fig:lli7-wegner} as a function of the flow parameter $\alpha_Y$.
The behavior of the energies is very similar to the standard generator and the (missing) induced YNN interactions are of similar or even larger size.
At the same time the overbinding due to the omitted induced YNN contributions develops in exactly the same way as the suppression of the $\Sigma$ admixture in the eigenstates.

The mechanism behind this is the following: consider a $\Lambda$ that is converted to a $\Sigma$ via interaction with a nucleon and then converted back by interaction with a different nucleon.
In a scheme with explicit $\Sigma$ hyperons, this process is an iterated two-body interaction. However, in a scheme without $\Sigma$ degrees of freedom this process appears as a genuine three-body $\Lambda$NN force.
The SRG evolution continuously transforms the problem from a $\Sigma$-full to a $\Sigma$-less scheme, i.e., it integrates out $\Sigma$ degrees of freedom, naturally generating those $\Lambda$NN interactions.
The presence of a strong repulsive YNN force is, therefore, merely a consequence of the degrees of freedom used in the calculation.
A calculation with full $\Lambda$--$\Sigma$ coupling needs a much weaker YNN force (or none at all) in order to reproduce experimental data.
Furthermore, the SRG is an ideal tool predict the $\Lambda$NN interactions for a $\Sigma$-less scheme starting from the full coupled-channel problem.

\paragraph{Implications for the hyperon puzzle.}
Two of the methods that are being used for microscopic calculations of strange neutron-star matter are Brueckner-Hartree-Fock and Auxiliary-Field Diffusion Monte Carlo (AFDMC).
Brueckner-Hartree-Fock calculations (see, e.g., Refs.\ \cite{Baldo2000,Takatsuka2002,Dapo2008,Vidana2000,Vidana2001,Vidana2011,Vidana2015,Yamamoto2013,*Yamamoto2014}) renormalize the interaction through a $G$\nobreakdash-matrix calculation that is truncated at the two-body level.
Similar to an SRG transformation at the two-baryon level, induced many-body terms are not captured in a $G$\nobreakdash-matrix framework and need to be accounted for by phenomenological corrections.

Recent AFDMC calculations by the Trento group \cite{Lonardoni2013,Lonardoni2014,Lonardoni2015} explicitly show the importance of $\Lambda$NN interactions for the equation of state.
Their present framework does not include $\Sigma$ hyperons and the $\Lambda$--$\Sigma$ conversion terms.
However, they fit phenomenological $\Lambda$N and $\Lambda$NN interactions to $\Lambda$ separation energies of various hypernuclei obtained in the same computational framework.
By comparing our $\Lambda$ separation energies for the $A=7$ and $A=13$ systems to the AFDMC results reported in Ref.\ \cite{Lonardoni2014} we notice that our calculations with induced YNN terms lie within or very close to the band spanned by the AFDMC results for the two fits of the $\Lambda$NN interaction they investigate.
Their AFDMC results without $\Lambda$NN interactions are well outside of this band and similar to our IT-NCSM results with the SRG-evolved Hamiltonian without the induced YNN terms.
These observations are fully in line with our previous discussion: The AFDMC without explicit $\Sigma$ degrees of freedom needs strongly repulsive phenomenological $\Lambda$NN interactions to reproduce the ground states of hypernuclei.
Our investigation shows that these strong $\Lambda$NN interactions necessarily emerge from the decoupling of the $\Sigma$ hyperons and originate from the $\Lambda$--$\Sigma$ conversion.
The AFDMC calculations also show the impact of the $\Lambda$NN interactions on the equation of state and on the maximum neutron-star mass.
Their inclusion stiffens the equation of state up to a point where the appearance of hyperons in the relevant density range is completely suppressed and the maximum neutron-star mass is set by the nucleonic equation of state.

\begin{table}
  %%% BLambda errors are estimated through fits to [NNax-8,NMax], [NNax-6,NMax] and [NNax-8,NMax-2].
  %%% Fits are repeated for different kappa_min extrapolations: using every kappa, only kappa <= 7e-5 for the hypernucleus and kappa <= 7e-5 for both.
  %%% The fit for Be8 uses Nmax=12 data not shown in the plot.
  \sisetup{table-format = 2.2(2)}
  \begin{tabular}{rSSSSS[table-space-text-post={[12]}]}
    \doubletoprule
    & {YN}       & \multicolumn{2}{c}{YN + ind. YNN} &    {AFDMC}    & \\
    \cmidrule{3-4}
    & {700}             & {600} & {700}   & {\cite{Lonardoni2013}} & {Experiment} \\
    \midrule
    \isotope[4][\Lambda]{He} &  4.10(1)  & 2.63(3)  & 2.56(4) & 1.22(9) & 2.39(3) \cite{Davis2005} \\
    \isotope[7][\Lambda]{He} &  9.93(36) & 7.41(34) & 5.98(33) & 5.95(25) & 5.68(28) \cite{Nakamura2013} \\
    \isotope[7][\Lambda]{Li} & 10.49(16) & 7.70(16) & 6.40(16) & & 5.58(3) \cite{Davis2005} \\
    \isotope[9][\Lambda]{Be} & 14.06(30) & 10.41(29) & 8.45(29) & & 6.71(4) \cite{Davis2005} \\
    \isotope[13][\Lambda]{C} & 20.06(10) & 17.50(21) & 14.43(19) & 11.2(4)& 11.69(12) \cite{Davis2005} \\
    \botrule
  \end{tabular}
  \caption{\label{tab:blambda}Extrapolated hyperon separation energies $B_\Lambda$ in \si{\MeV} for selected hypernuclei.
  The SRG flow parameter is $\alpha_N=\alpha_Y=\SI{0.08}{\fm\tothe4}$, the HO frequency $\hbar\Omega=\SI{20}{\MeV}$.
  Numbers in header rows denote the YN interaction cutoff $\Lambda$ in $\si{\MeVc}$.}
\end{table}

\paragraph{Conclusions.}
We present the first \emph{ab initio} calculations of $p$\nobreakdash-shell hypernuclei with an explicit treatment of YNN interactions induced by SRG transformations.
The inclusion of the induced YNN terms removes the overbinding and the flow-parameter dependence of low-lying states in light hypernuclei, providing a good description of hypernuclear spectroscopy already with LO chiral YN interactions.
We demonstrate that the SRG-induced YNN terms are driven by the suppression of the $\Lambda$--$\Sigma$ coupling in the YN interaction, which promotes certain iterated two-body interactions to the three-body level.
Effectively the SRG provides a continuous mapping between a scheme with fully coupled $\Lambda$ and $\Sigma$ degrees of freedom and a simplified scheme with only $\Lambda$ hyperons at the expense of strong repulsive $\Lambda$NN interactions.
Our findings explain why phenomenological models using only $\Lambda$ hyperons necessarily need strong $\Lambda$NN interactions or corresponding density dependencies to give realistic results.
This also provides evidence that the $\Lambda$--$\Sigma$ conversion in a fully coupled theory is key to resolving the hyperon puzzle even without the use of additional chiral YNN interactions.

\begin{acknowledgments}
The authors gratefully acknowledge the computing time granted on the supercomputers JURECA at the Jülich Supercomputing Centre, LOEWE at the Center for Scientific Computing Frankfurt, and Lichtenberg at Technische Universit\"at Darmstadt.
This work is supported by the Deutsche Forschungsgemeinschaft through grant SFB 1245, the Helmholtz International Center for FAIR, and the BMBF through Contract No.\ 05P15RDFN1.
\end{acknowledgments}

\end{document}